\documentclass[preprint,showpacs,preprintnumbers,amsmath,amssymb]{revtex4}
\usepackage{graphicx,epsfig,subfigure,latexsym,makeidx,amsmath}
\usepackage{color}
\usepackage{float}
\usepackage[colorlinks,linkcolor=blue,anchorcolor=blue,citecolor=green,urlcolor=blue]{hyperref}
\usepackage[english]{babel}

\begin{document}
	%%%%%%%%%%%%%%%%%%%%%%%%%%%%%%%%%%%%%%%%%%%%%%%%%%%%%%%%%%%%%%%%%%%%%%%%%%%%%%%%%%%%%%
	\newcommand \nn{\nonumber}
	\newcommand \fc{\frac}
	\newcommand \lt{\left}
	\newcommand \rt{\right}
	\newcommand \pd{\partial}
	\newcommand \e{\text{e}}
	\newcommand \hmn{h_{\mu\nu}}
	\newcommand{\PR}[1]{\ensuremath{\left[#1\right]}} % parenteses rectos do tamanho adequado
	\newcommand{\PC}[1]{\ensuremath{\left(#1\right)}} % parenteses curvos do tamanho adequado
	\newcommand{\PX}[1]{\ensuremath{\left\lbrace#1\right\rbrace}} % chavetas do tamanho adequado
	\newcommand{\BR}[1]{\ensuremath{\left\langle#1\right\vert}} % Bra do tamanho adequado
	\newcommand{\KT}[1]{\ensuremath{\left\vert#1\right\rangle}} % Ket do tamanho adequado
	\newcommand{\MD}[1]{\ensuremath{\left\vert#1\right\vert}} % modulo do tamanho adequado
    \newcommand{\tcr}{\textcolor{red}}
    \newcommand{\tcb}{\textcolor{blue}}
	%%%%%%%%%%%%%%%%%%%%%%%%%%%%%%%%%%%%%%%%%%%%%%%%%%%%%%%%%%%%%%%%%%%%%%%%%%%%%%%%%%%
	
	\title{Charged Spherically Symmetric and Slowly Rotating Charged Black Hole  Solutions in Bumblebee Gravity }
	
	\preprint{}

	\author{Jia-Zhou Liu$^1$ $ ^2$, Wen-Di Guo$^1$ $ ^2$, Shao-Wen Wei$^1$ $^2$, and Yu-Xiao Liu$^1$ $^2$ \footnote{Corresponding author. E-mail: liuyx@lzu.edu.cn}}

	\affiliation{$^{1}$Lanzhou Center for Theoretical Physics, Key Laboratory of Theoretical Physics of Gansu Province,  Key Laboratory for Quantum Theory and Applications of the Ministry of Education, Gansu Provincial ResearchCenter for Basic Disciplines of Quantum Physics,  Lanzhou University, Lanzhou 730000, China,\\
		$^{2}$Institute of Theoretical Physics $\&$ Research Center of Gravitation, School of Physical Science and Technology, Lanzhou University, Lanzhou 730000, China}

	\begin{abstract}
In this paper, we study the scenario in which the matter field is an electromagnetic field nonminimally coupled to the bumblebee vector field. We present exact charged spherically symmetric black hole solutions and slowly rotating charged solutions in bumblebee gravity with and without a cosmological constant. The static spherically symmetric solutions describe the Reissner-Nordström-like black hole and Reissner-Nordström-(anti) de Sitter-like black hole, while the stationary and axially symmetric soltuions describe the Kerr-Newman-like black hole and Kerr-Newman-(anti) de Sitter-like black hole. We utilize the Hamilton-Jacobi formalism to study the shadows of the black holes. Additionally, we investigate the effect of the electric charge and Lorentz-violating parameters on the radius of the shadow reference circle and the distortion parameter. We find that the radius of the reference circle decreases with the Lorentz-violating parameter and the charge parameter, while the distortion parameter increases with the Lorentz-violating parameter and the charge parameter.	
	
\end{abstract}	

\maketitle
\newpage

\newpage

\section{INTRODUCTION}\label{intro}

In the area of classical physics, gravitational phenomena are well-explained by general relativity, which has endured rigorous experimental and theoretical validation. On the quantum front, the Standard Model of particle physics has been successfully described the other three fundamental interactions. These two theories offer a comprehensive and understanding of the natural world. Therefore, the quest to unify these two theories is fundamental, and its achievement will inevitably lead us to a deeper comprehension of nature.

Several quantum gravity theories have been proposed~\cite{Birrell:1982ix,Maldacena:1997re,Aharony:1999ti,Gubser:1998bc,Alfaro:1999wd,Alfaro:2001rb,Rovelli:1989za}, but it is needed the experiments at the Planck scale ($\sim 10^{19}$ GeV) to observe the quantum effect, this is beyond the ability of current experiments. However, some  quantum gravity theories assume that Lorentz symmetry might be broken in the gravitational UV regime which is at our presently accessible low-energy scales. Stemming from the concept of spontaneous Lorentz symmetry breaking (LSB) in string theory~\cite{Kostelecky:1988zi,Kostelecky:2003fs}, the Standard Model extension  was introduced as a suitable framework for violating Lorentz symmetry~\cite{Colladay:1996iz,Colladay:1998fq,Kostelecky:2003fs}.In the following decades, the Standard Model extension has been widely studied~\cite{Colladay:2001wk,Kostelecky:2000mm,Kostelecky:2001mb,Colladay:2009rb,Berger:2015yha,Carroll:1989vb,Andrianov:1994qv,Andrianov:1998wj,Lehnert:2004hq,BaetaScarpelli:2012kt,Brito:2013npa}.

In 1989, Kostelecky and Samuel introduced the prototype of the bumblebee model~\cite{Kostelecky:1989jw,Kostelecky:1988zi}, a string-inspired framework featuring tensor-induced spontaneous Lorentz symmetry breaking. In the framework of bumblebee gravity, the  spontaneous Lorentz violation arises from a potential $V(B^\mu B_\mu)$ acting on an vector field $B_\mu$~\cite{Kostelecky:2003fs,Kostelecky:2010ze}. Not just in Minkowski's spacetime, the bumblebee model can also be explored in  Riemann and  Riemann-Cartan spacetimes~\cite{Bailey:2006fd,Maluf:2013nva,Li:2020dln,Bluhm:2008yt,Escobar:2017fdi,Maluf:2015hda,Hernaski:2014jsa,Carroll:2009em}.

The pursuit of black hole solutions holds significance for any theory of gravity. In 2017, Casana et al. obtained an exact Schwarzschild-like black hole solution in the bumblebee gravity~\cite{Casana:2017jkc}. Subsequently, researchers have independently proposed exact solutions for traversable wormholes~\cite{Ovgun:2018xys}, Schwarzschild-anti-de Sitter (AdS)-like black holes~\cite{Maluf:2020kgf}, slowly rotating Kerr-like black holes~\cite{Ding:2019mal}, and other solutions~\cite{Santos:2014nxm,Jha:2020pvk,Filho:2022yrk,Xu:2022frb,Ding:2023niy} within bumblebee gravity. Moreover, in other scenarios within the framework of the Standard Model Extension, such as when considering the nonminimally coupled Kalb-Ramond field with a nonzero vacuum expectation value, certain black hole solutions have also been obtained~\cite{Lessa:2019bgi,Kumar:2020hgm,Yang:2023wtu,Duan:2023gng,Do:2020ojg,Liu:2024oas}. And the properties of the blakc holes were also studied~\cite{Guo:2023nkd,Du:2024uhd,Kuang:2022xjp,Oliveira:2018oha,Liu:2022dcn,Gomes:2018oyd,Kanzi:2019gtu,Gullu:2020qzu,Oliveira:2021abg,Kanzi:2021cbg,Hosseinifar:2024wwe,Liu:2024lve,Liu:2024oas,Ovgun:2018ran,Sakalli:2023pgn,Mangut:2023oxa,Uniyal:2022xnq}.

In 1966, Synge initiated the examination of the shadow cast by a  Schwarzschild black hole~\cite{Synge:1966okc}. Subsequently, Luminet~\cite{Luminet:1979nyg} studied the influence of a thin accretion disk on the black hole shadow, while Bardeen~\cite{Bardeen:1973tla} investigated the shadow of a Kerr black hole. Over the decades, significant progress has been made in the study of black hole shadows~\cite{deVries:1999tiy,Takahashi:2005hy,Bambi:2011yz,Atamurotov:2013sca,Perlick:2018iye,Meng:2023uws,Sui:2023rfh}. Recently, direct imaging of the black holes M87* at the core of the Virgo A galaxy and Sgr A* at the core of the Milky Way galaxy has been achieved  by the sub-millimeter Event Horizon Telescope (EHT) employing very-long baseline interferometry~\cite{EventHorizonTelescope:2019dse,EventHorizonTelescope:2019ggy,EventHorizonTelescope:2019pgp,EventHorizonTelescope:2019ths,EventHorizonTelescope:2022wkp}. Therefore, investigating the black hole shadow is highly desirable.

Our aim in this paper is to obtain Charged spherically symmetric and slowly rotating charged black
hole solutions. Subsequently, using the equations governing the motion of photons, we analyze  the shadows cast by rotating black holes, exploring the impact of Lorentz violation on the size and deformation of these shadows. Calculations are then conducted to determine the radius and distortion observables.

This paper is structured as follows: Sec.~II provides a review of the Einstein-Bumblebee theory. In Sec.~III, we present the charged bumblebee black hole solution in both the presence and absence of a cosmological constant. And in Sec.~IV, we extend this solution to the case of rotating black holes. Sec.~V is devoted to investigating the shadow of rotating black holes and computing relevant parameters. Finally, Sec.~VI presents a summary and discussion of the work.

\section{ EINSTEIN-BUMBLEBEE THEORY }%\label{spacetimes}
As discussed in the preceding section, the bumblebee model extends general relativity by adding a vector field called bumblebee field which couples with gravity nonminimally. The bumblebee vector field $B_{\mu}$ obtains a nonzero vacuum expectation value through a specified potential, resulting in the spontaneous breaking of Lorentz symmetry in the gravitational sector. The action is described by ~\cite{Kostelecky:2003fs}

\begin{eqnarray}
	S&=&\int d^{4}x\sqrt{-g} \bigg [  \frac{1}{2\kappa}\left(R-2\Lambda\right)+\frac{\xi}{2\kappa}B^{\mu}B^{\nu}R_{\mu\nu}-\frac{1}{4} B_{\mu\nu}B^{\mu\nu}-V(B^{\mu}B_{\mu}\pm b^{2}) \bigg ] \nonumber \\
	&&+\int d^{4}x\sqrt{-g}\mathcal{L}_{M}, 
	\label{S}
\end{eqnarray}
where $\Lambda$ denotes the cosmological constant,  $\kappa=\frac{8\pi G}{c^4}$ is the gravitational coupling constant, and $\xi$ represents the nonminimal coupling constant between gravity and the bumblebee field. Analogous to the electromagnetic field, the bumblebee field strength is given by
\begin{equation}
	B_{\mu\nu}=\partial_{\mu}B_\nu-\partial_{\nu}B_{\mu}.
	\label{BB}
\end{equation}
The potential $V$ provides a nonvanishing vacuum expectation value for $B_\mu$ and could take the functional form of $V(B^{\mu}B_{\mu}\pm b^{2})$, implying that the condition $B^{\mu}B_{\mu}=\mp b^{2}$ must be satisfied. In essence, the field $B_{\mu}$ acquires a nonvanishing vacuum expectation value $\left\langle B_{\mu}\right\rangle = b_{\mu}$, where the vector $b_{\mu}$ is a function of the spacetime coordinates and satisfies $b^{\mu}b_{\mu}=\mp b^{2}=\mathrm{const}$. For convenience, we define
$X=B^{\mu}B_{\mu}\pm b^{2}$ and
$V'=\frac{\partial V}{\partial X}$, which will be used in the subsequent discussions.

We consider the matter field to be an electromagnetic field  nonminimally coupled with the bumblebee vector field~\cite{Lehum:2024ovo}. Its Lagrangian density expression reads as
\begin{eqnarray}
	\mathcal{L}_{M}&=&\frac{1}{2\kappa}(F^{\mu\nu}F_{\mu\nu}+{\gamma}B^{\mu}B_{\mu}F^{\alpha\beta}F_{\alpha\beta}),
	\label{L}
\end{eqnarray} 
where the electromagnetic tensor related to the electromagnetic field is

\begin{eqnarray}
	&&F_{\mu\nu}=\partial_{\mu}A_\nu-\partial_{\nu}A_{\mu},\nonumber\\
	&&   A =(\phi(r),0,0,0) ,
	\label{Fa}
\end{eqnarray}
and ${\gamma}$ presents the coupling coefficient. The gravitational field equations in the framework of bumblebee gravity can be derived by varying the action \eqref{S} with respect to the metric tensor $g^{\mu\nu}$:
\begin{equation}
	G_{\mu\nu}+\Lambda g_{\mu\nu}= \kappa T^{B}_{\mu\nu}+ \kappa T^{M}_{\mu\nu} ,
	\label{modified}
\end{equation}
where
\begin{eqnarray}
	T^{B}_{\mu\nu}&=&
	 \frac{\xi}{\kappa}\left[\frac{1}{2}B^{\alpha}B^{\beta}R_{\alpha\beta}g_{\mu\nu}-B_{\mu}B^{\alpha}R_{\alpha\nu}-B_{\nu}B^{\alpha}R_{\alpha\mu}\right.+\frac{1}{2}\nabla_{\alpha}\nabla_{\mu}\left(B^{\alpha}B_{\nu}\right)\nonumber\\
	&&  +\frac{1}{2}\nabla_{\alpha}\nabla_{\nu}\left(B^{\alpha}B_{\mu}\right)\left.-\frac{1}{2}\nabla^{2}\left(B_{\mu}B_{\nu}\right)-\frac{1}{2}
	g_{\mu\nu}\nabla_{\alpha}\nabla_{\beta}\left(B^{\alpha}B^{\beta}\right)\right] \nonumber\\
	&&+2V'B_{\mu}B_{\nu} +B_{\mu}^{\ \alpha}B_{\nu\alpha}-\left(V+ \frac{1}{4}B_{\alpha\beta}B^{\alpha\beta}\right)g_{\mu\nu},
	\label{fa}
\end{eqnarray}
and
\begin{equation}
	T^{M}_{\mu\nu}=\frac{1}{\kappa}\left[(1+\gamma b^2)(2F_{\mu\alpha}F_{\nu}^{\alpha}-\frac{1}{2}g_{\mu\nu}F^{\alpha\beta}F_{\alpha\beta})+\gamma B_{\mu}B_{\nu}F^{\alpha\beta}F_{\alpha\beta}\right].
	\label{Tm}
\end{equation}
For computational convenience, we can express the field equations in the following form:
\begin{equation}
R_{\mu\nu}=\Lambda g_{\mu\nu}+ \kappa	\mathcal{T}_{\mu\nu} ,
	\label{modified}
\end{equation}
where
\begin{eqnarray}
	\mathcal{T}^{M}_{\mu\nu}&=&{T}^{M}_{\mu\nu}-\frac{1}{2}{T}^{M}g_{\mu\nu}\nonumber\\
	\mathcal{T}^{B}_{\mu\nu}&=&{T}^{B}_{\mu\nu}-\frac{1}{2}{T}^{B}g_{\mu\nu}
	\label{RR}
\end{eqnarray}
and

\begin{eqnarray}
	\kappa\mathcal{T}_{\mu\nu}&=&\kappa\mathcal{T}^{M}_{\mu\nu}+\kappa\mathcal{T}^{B}_{\mu\nu}\nonumber\\
	&=&\kappa\left(T^M_{\mu\nu}-\frac{1}{2}g_{\mu\nu}T^M\right)+  \kappa\left[V'\left( 2 B_{\mu}B_{\nu}-b^{2}g_{\mu\nu}\right) +B_{\mu}^{\ \alpha}B_{\nu\alpha}+V g_{\mu\nu}- \frac{1}{4}B_{\alpha\beta}B^{\alpha\beta}g_{\mu\nu} \right]\nonumber\\&&+\xi\left[\frac{1}{2}B^{\alpha}B^{\beta}R_{\alpha\beta}g_{\mu\nu}-B_{\mu}B^{\alpha}R_{\alpha\nu}-B_{\nu}B^{\alpha}R_{\alpha\mu}+\frac{1}{2}\nabla_{\alpha}\nabla_{\mu}\left(B^{\alpha}B_{\nu}\right)\right. +\frac{1}{2}\nabla_{\alpha}\nabla_{\nu}\left(B^{\alpha}B_{\mu}\right)\nonumber\\
	&&\left.-\frac{1}{2}\nabla^{2}\left(B_{\mu}B_{\nu}\right)\right] .
	\label{RR}
\end{eqnarray}
Similarly, by varying the action \eqref{S} with respect to the bumblebee vector field and the electromagnetic field, we can obtain the equations of motion for the corresponding  fields:
\begin{eqnarray}
	\nabla_{\mu}B^{\mu\nu}-2\left( V'B^{\nu}-\frac{\xi}{2\kappa}B_\mu R^{\mu\nu}-\frac{1}{2\kappa}\gamma B^{\nu}F^{\alpha\beta}F_{\alpha\beta} \right)&=&0,\label{BB}\\ 	
		\nabla_{\mu}\left(F^{\mu\nu}+\gamma B^{\alpha}B_{\alpha}F^{\mu\nu}\right)&=&0.
	\label{FF}
\end{eqnarray}

\section{CHARGED SPHERICALLY SYMMETRIC BLACK HOLE SOLUTIONS }
We consider the metric ansatz for a static and spherically symmetric spacetime, which is expressed as
\begin{equation}
	{d}{s}^{2}=-A(r) {dt}^{2}+S(r
	){dr}^{2}+r^{2} {~d}\Omega ^{2},
	\label{qdc}
\end{equation}
where ${~d} \Omega ^{2}={~d} \theta^{2}+ \sin ^{2} \theta {d} \varphi^{2}$.

In the previous study conducted by Casana et al., they formulated an exact black hole solution without a cosmological constant and the matter field 
$A_{\mu}$~\cite{Casana:2017jkc}. In the scenario where the bumblebee field $B_\mu$ maintains its vacuum expectation value $b_\mu$, the bumblebee field is given by~\cite{Bertolami:2005bh}
\begin{equation}B_{\mu}=b_{\mu}.\end{equation}
Similar to Ref.~\cite{Casana:2017jkc}, we consider a spacelike background $b_\mu$ with the form
\begin{equation}b_{\mu}=\left(0,b_{r}(r),0,0\right).\end{equation}
Utilizing the aforementioned condition $b_{\mu}b^{\mu}=b^{2}=const$, we can derive
\begin{equation}b_r(r)=b\sqrt{S(r)}.\end{equation}

For convenience, we set $\ell=\xi b^2$ as  Lorentz-violating
parameter. By combining the spherically symmetric metric, we can obtain the specific field equations:

\begin{eqnarray}
	\frac{(2+\ell) A'(r)^2}{8 A(r) S(r)}-\frac{(2+\ell) A'(r) S'(r)}{8 S(r)^2}-\frac{(1+\ell) A'(r)}{r S(r)}+\frac{(2+\ell) A''(r)}{4 S(r)}&&\\
	-\frac{\ell A(r) S'(r)}{2 r S(r)^2}-(1+2 b^2 \gamma)A(r)\phi'(r)^2+\Lambda A(r) +\kappa(V'b^2 +V) A(r)&=&0,\label{G1} \\
	\frac{A'(r) S'(r)}{4 A(r) S(r)}-\frac{A''(r)}{2 A(r)}+\frac{A'(r)^2}{4 A(r)^2}+\frac{S'(r)}{r S(r)}&&\nonumber
	\\+\frac{2 (1+2 b^2 \gamma)\phi'(r)^2 S(r)}{(2+3\ell) }-\frac{2\Lambda S(r)}{(2+3\ell)}-2\kappa \frac{(V'b^2 +V) S(r)}{(2+3\ell)}&=&0,\label{G2} \\
	1-\frac{1+\ell}{S(r)}+\frac{r S'(r)}{2 S(r)^2}+\frac{\ell r^2 A''(r)}{4 A(r) S(r)}-\frac{\ell r^2 A'(r) S'(r)}{8 A(r) S(r)^2}-\frac{\ell r^2 A'(r)^2}{8 A(r)^2 S(r)}&&\nonumber\\
	-\frac{(1+\ell) r A'(r)}{2 A(r) S(r)}-(1+2 b^2 \gamma)  \phi'(r)^2r^2-\Lambda r^2-\kappa(V'b^2 +V)r^2&=&0,\label{G3}\\
	\frac{ A''(r)}{2 A(r)}-\frac{ A'(r)^2}{4 A(r)^2}-\frac{ A'(r) S'(r)}{4 A(r) S(r)}-\frac{2\gamma \phi'(r)^2}{\xi \left(1+\ell\right)A(r)}-\frac{ S'(r)}{r S(r)}+\frac{2 \kappa  }{\xi }S(r) V'&=&0,\label{bb} \\
	\frac{d}{dr}\left[r^2\frac{F_{tr}}{\sqrt{A(r)S(r)}}\right]&=&0.
	\label{fff}
\end{eqnarray}

Next, we will solve the above 
equations and give spherically symmetric black hole solutions under different scenarios.
\subsection{ Case A: $V(X)=\frac{\lambda}{2}X^2$ and $ \Lambda=0$ }

In the absence of the cosmological constant, similar to the work of Casana et al., we impose the vacuum conditions  $V=0$ and $V'=0$~\cite{Casana:2017jkc}.
An illustrative example of the potential satisfying these conditions is readily presented by a smooth quadratic form:
\begin{equation}
	V(X)=\frac{\lambda}{2}X^2,
	\label{VV}
\end{equation}
 where $\lambda$ is a constant.This form is same as the  potential form of the Higgs field, and is also related to the mass structure of the theory~\cite{Kostelecky:1989jw}. In this case, the potential $V$ has no contribution to the field equations. Other choices of the potential, such as $V(X) \equiv \frac{\lambda}{2} X^n$ ($n\geq3$), also have no contribution to the field equations. Therefore, the solutions for these choices are consistent with the potential $ V(X) = \frac{\lambda}{2} X^2 $.

Using Eq.~\eqref{fff}, we can derive $F_{tr}=\sqrt{A(r)S(r)}\phi'(r)$. Substituting the expressions of 
$F_{tr}$ here and  
$V$ in Eq.~\eqref{VV} into above conditions into the field equations~\eqref{G1} - \eqref{fff}, and considering the value of the coupling parameter $\gamma$ to be $\frac{\xi}{2+\ell}$, the field equations are given by:

\begin{eqnarray}
	\frac{(2+\ell) A'(r)^2}{8 A(r) S(r)}-\frac{(2+\ell) A'(r) S'(r)}{8 S(r)^2}-\frac{(1+\ell) A'(r)}{r S(r)}&& \nonumber\\
	+\frac{(2+\ell) A''(r)}{4 S(r)}-\frac{\ell A(r) S'(r)}{2 r S(r)^2}-\frac{A(r)\phi'(r)^2}{ \left(1+\ell\right)}&=&0,\label{S1}\\
	\frac{A'(r) S'(r)}{4 A(r) S(r)}-\frac{A''(r)}{2 A(r)}+\frac{A'(r)^2}{4 A(r)^2}+\frac{S'(r)}{r S(r)}+\frac{2 \phi'(r)^2 S(r)}{(2+\ell)}&=&0,\label{S2} \\ 
	1-\frac{1+\ell}{S(r)}+\frac{r S'(r)}{2 S(r)^2}+\frac{\ell r^2 A''(r)}{4 A(r) S(r)}-\frac{\ell r^2 A'(r) S'(r)}{8 A(r) S(r)^2}-\frac{\ell r^2 A'(r)^2}{8 A(r)^2 S(r)}&& \nonumber\\
	-\frac{(1+\ell) r A'(r)}{2 A(r) S(r)}-\frac{(2+3\ell) \phi'(r)^2r^2}{(2+\ell)}&=&0,\\ \nonumber\\
	\frac{ A''(r)}{2 A(r)}-\frac{ A'(r)^2}{4 A(r)^2}-\frac{ A'(r) S'(r)}{4 A(r) S(r)}-\frac{\phi'(r)^2}{ \left(1+\ell\right)A(r)}-\frac{ S'(r)}{r S(r)}&=&0, \\ \nonumber\\
	\frac{d}{dr}[r^2\phi'(r)]&=&0.
	\label{33}
\end{eqnarray}
By simplifying  Eqs.~\eqref{S1} and \eqref{S2}, we obtain:
\begin{equation}
	[S(r) A(r)]'=0,
\end{equation}
which implies that $S(r)=\frac{C_1}{A(r)}$, where $C_1$ is a constant. Similar to the Schwarzschild-like black hole solution in bumblebee gravity, we set the value of $C_1$ to be $1+\ell$~\cite{Casana:2017jkc}. Building upon this, we solve the modified  Maxwell's equations to obtain:
$\phi(r)=\frac{C_2}{r}+C_3$. Here, we set $C_2=Q_{0}$ and $C_3=0$ (i.e., $\phi(r)=\frac{Q_{0}}{r}$), and we ultimately arrive at the final solution:
\begin{eqnarray}
	A(r)&=&1-\frac{2 M}{r}+\frac{ 2(1+\ell)Q_0^2}{(2+\ell) r^2},\label{Q11}\\
	S(r)&=&\frac{1+\ell}{A(r)},\label{Q12}\\
	\phi(r)&=&\frac{Q_{0}}{r}.
\end{eqnarray}

Through the modified Maxwell's equations, the conserved current is modified to be $J^{\nu}=\nabla_{\mu}\left(F^{\mu\nu}+\gamma B^{\alpha}B_{\alpha}F^{\mu\nu}\right)$. Then, we can obtain the integration constants, the  Maxwell's equations was also modified. 

\begin{eqnarray}
	Q&=&-\frac{1}{4\pi}\int_{\Sigma}dx^{3}\sqrt{\gamma^{(3)}}n_{\mu}J^{\mu} \nonumber\\
	&=&-\frac{1}{4\pi}\int_{\partial\Sigma}d\theta d\phi\sqrt{\gamma^{(2)}}n_{\mu}\sigma_{\nu}\left(F^{\mu\nu}+\frac{\xi }{l +2} B^{\alpha}B_{\alpha}F^{\mu\nu}\right) \nonumber\\
	&=&\left(1+b^{2}\frac{\xi }{(l +2)}\right)Q_0 \nonumber\\	
	&=&\frac{2(1+\ell)}{2+\ell}Q_0.
\end{eqnarray}
Here $\Sigma$ represents a three-dimensional spacelike region with the induced metric $\gamma_{ij}^{(3)}$, the  boundary $\partial\Sigma$ is a two-sphere located at spatial infinity with the induced metric $\gamma_{ij}^{(2)}=r^{2} {~d}\Omega ^{2}$, while $n_\mu=(1,0,0,0)$ and $\sigma_{\mu}=(0,1,0,0)$ denote the unit normal vectors 
associated with $\partial\Sigma$ and $\Sigma$, respectively.  
The solution of $A(r)$ can also be rewritten as
\begin{equation}
	A(r)=1-\frac{2 M}{r}+\frac{(2+\ell) Q^2}{2(1+\ell) r^2},
\end{equation}
this solution is similar to the Reissner-Nordström (RN) solution. As $\ell$ tends to 0, it recovers to the RN solution.  The introduction of Lorentz symmetry breaking distinguishes it from  many modified gravity theories. It is  evident that the metric functions approach to $A(r)\rightarrow1$ and $S(r)\rightarrow1+\ell$ at infinity, indicating that the  spacetime is not asymptotically Minkowski.Furthermore, the $ g_{tt} $ component of the  metric \eqref{qdc} differs from the RN black hole solution only by the modification of the coefficient of the $ \frac{Q^2}{r^2} $ term.  However, the electrically charged black hole solutions in many modified gravity theories modify the terms involving both $ \frac{M}{r} $ and $ \frac{Q^2}{r^2} $ in the RN metric, or introduce new terms. For example, the charged black hole solutions in Einstein-Gauss-Bonnet gravity~\cite{Fernandes:2020rpa} and Eddington-inspired Born–Infeld gravity~\cite{Wei:2014dka} modify the terms involving both $ \frac{M}{r} $ and $ \frac{Q^2}{r^2} $, while the charged black hole solution in Einstein-Maxwell-aether gravity introduces new term~\cite{Ding:2015kba}.  In the paper~\cite{Casana:2017jkc}, Casana et al. studied the effects of spontaneous Lorentz symmetry breaking and investigated some classic tests, including the advance of Mercury’s perihelion, bending of light, and Shapiro’s time-delay. Furthermore, they have computed several upper bounds, with the most stringent sensitivity of $\ell<6.2\times10^{-13}$.

The introduction of the nonminimal
	coupling parameter $\ell$ does have some effects on black hole thermodynamics, such as the black hole temperature, entropy, and so on. However, it does not fundamentally alter the number of possible black hole phases, the phase transition behaviors, and the thermodynamic stability. In Appendix A, we have provided a series of specific analyses and discussions.
 	
 Similar to the RN black hole, this solution exhibits two horizons:
\begin{equation}
	r_{\pm}=M \pm \sqrt{M^2-\frac{2 (\ell+1) }{\ell+2}Q_0^2} .
\end{equation}
It is obvious that  the mass and charge parameter of the black hole should satisfy
\begin{equation}
	\frac{ Q_0^2}{M^2}\leq \frac{(2+\ell)} {2(1+\ell)}.
\end{equation}
Another aspect of this solution is the physical singularity which can be studied through the Kretschmann scalar  
$K$, derived from the Riemann tensor:
\begin{eqnarray}
	K&=&R_{\alpha\beta\delta\gamma}R^{\alpha\beta\delta\gamma}\\
	&=&\frac{4 \left(\ell^2 r^2+4 l M r+12 M^2\right)}{(\ell+1)^2 r^6}-\frac{16 Q_0^2 \left((\ell+2) r (\ell r+12 M)-14 (\ell+1) Q_0^2\right)}{(\ell+1) (\ell+2)^2 r^8}.
	\end{eqnarray}
From this we can see that, the singularity only appears at $r=0$ (assuming 
$0<\left|\ell\right|\ll 1$) which is similar to  the singularity of the RN black hole. 

To further clarify and describe the spacetime structure of the  Charged black hole in Bumblebee gravity, we will present the Penrose diagram. By performing a coordinate transformation, we obtain:
\begin{equation}
	{d}{s}^{2}=-(1-R_+)(1-R_-) {dt}^{2}+\frac{1}{(1-R_+)(1-R_-)}{dR}^{2}+\frac{1}{(1+\ell)}R^{2} {~d}\Omega ^{2},
	\label{pd11}
\end{equation}
\begin{figure}
	\begin{center}
		\includegraphics[width=5cm]{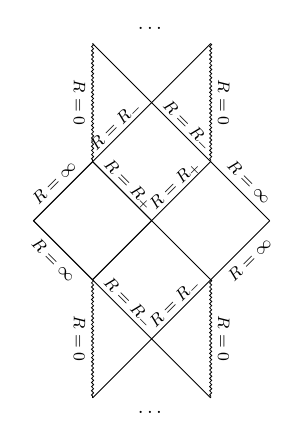}
	\end{center}
	\caption{ Penrose diagram of the non-extremal charged black hole in bumblebee gravity.}
	\label{Penrose diagram1}
\end{figure}where $R$ is the parameter after coordinate transformation of $r$ ($r=\sqrt{\frac{1}{1+\ell}}R$) and $R_{\pm}$ are the result of the two horizon radii after the coordinate transformation ($R_{\pm}=\sqrt{\frac{1}{1+\ell}}(M \pm \sqrt{M^2-\frac{2 (\ell+1) }{\ell+2}Q_0^2})$). The corresponding Penrose diagrams are illustrated in Fig.~\ref{Penrose diagram1}. It is evident that the Penrose diagram of this solution is identical to that of the  RN spacetime.  The number and type of horizons of the electrically charged black holes obtained in our paper are indeed consistent with the  RN  black hole. The speed of light remains unchanged, and the causal structure is unaffected. However, due to  Lorentz symmetry breaking in bumblebee gravity, the asymptotic behavior of the resulting spacetime is not Minkowskian. In addition, although the Penrose diagram obtained here is the same as that of the  RN black hole, the two-dimensional spheres at the same points on the Penrose diagrams are not the same. Two horizon radii of the electrically charged black hole in bumblebee gravity differ from those of the RN black hole, and there is an additional correction factor $ \frac{1}{(1+\ell)} $ in front of the $ R^2 d\Omega^2 $ term compared to the RN black hole. It follows that, although our Penrose diagram is identical to that of the RN black hole   after a series of coordinate transformations, the radii of the corresponding two-dimensional spheres in the diagrams are different.

\subsection{Case B: $V(X)=\frac{\lambda}{2}X$ and $ \Lambda\neq0$}

Next, we consider the scenario involving a nonzero cosmological constant. Our purpose is to derive an analytical black hole solution in this context. The geometry of the solution is similar to the RN-AdS black hole or the RN-dS black hole. Based on  the groundwork by R.V. Maluf et al.~\cite{Maluf:2020kgf}, conditions are  different from those in the previous discussion. The straightforward option for the potential can be a linear function:

\begin{equation}
	V(X)=\frac{\lambda}{2}X,
\end{equation}
where $\lambda$ is a Lagrange multiplier~\cite{Bluhm:2007bd}. Note that the equation of motion for the Lagrange multiplier ensures the vacuum condition $X=0$, which also ensures that $V=0$ for any field $\lambda$ on-shell. However, unlike the vacuum condition in Case A, $V^{\prime}\neq0$ is valid when the field $\lambda$ is nonzero. Thus, the potential $V$ contributes to the field equations. The choice $V(X) = \frac{\lambda}{2} X$ effectively freezes motion about the potential minimum and hence permits an efficient extraction of the essential physics~\cite{Kostelecky:1989jw}. The constraint for the Lagrange multiplier $\lambda$ is providedthe field Eqs~\eqref{RR} and \eqref{BB} .

Combining the spherically symmetric metric and the form of the potential $V$, we can derive the specific field equations:
\begin{eqnarray}
	\frac{(2+\ell) A'(r)^2}{8 A(r) S(r)}-\frac{(2+\ell) A'(r) S'(r)}{8 S(r)^2}-\frac{(1+\ell) A'(r)}{r S(r)}&& \nonumber\\
	+\frac{(2+\ell) A''(r)}{4 S(r)}-\frac{\ell A(r) S'(r)}{2 r S(r)^2}-\frac{A(r)\phi'(r)^2}{ \left(1+ \ell\right)}+\frac{\Lambda  A(r) (2+\ell)}{2\ell}&=&0,\\
	\frac{A'(r) S'(r)}{4 A(r) S(r)}-\frac{A''(r)}{2 A(r)}+\frac{A'(r)^2}{4 A(r)^2}+\frac{S'(r)}{r S(r)}+\frac{2 \phi'(r)^2 S(r)}{(2+\ell)}-\frac{\Lambda S(r)}{1+\ell}&=&0,\\
	1-\frac{1+\ell}{S(r)}+\frac{r S'(r)}{2 S(r)^2}+\frac{\ell r^2 A''(r)}{4 A(r) S(r)}-\frac{\ell r^2 A'(r) S'(r)}{8 A(r) S(r)^2}-\frac{\ell r^2 A'(r)^2}{8 A(r)^2 S(r)}&& \nonumber\\
	-\frac{(1+\ell) r A'(r)}{2 A(r) S(r)}-\frac{(2+3\ell) \phi'(r)^2r^2}{(2+\ell)}-\frac{\Lambda  (2+\ell) r^2}{2 (1+\ell)}&=&0,\\
	\frac{ A''(r)}{2 A(r)}-\frac{ A'(r)^2}{4 A(r)^2}-\frac{ A'(r) S'(r)}{4 A(r) S(r)}-\frac{\phi'(r)^2}{ \left(1+\ell\right)A(r)}-\frac{ S'(r)}{r S(r)}+\frac{\Lambda S(r)}{1+\ell}&=&0, \\
	\frac{d}{dr}[r^2\phi'(r)]&=&0.
	\label{33}
\end{eqnarray}
An analytical solution is  possible if and only if
\begin{equation}
	\Lambda=\frac{\kappa\lambda}{\xi}(1+\ell).
\end{equation}
Under the above condition, we can obtain
\begin{eqnarray}
	A(r)&=&1-\frac{2 M}{r}+\frac{ 2(1+\ell)Q_0^2}{(2+\ell) r^2}-\frac{1}{3} (1+\ell)\Lambda _e r^2 ,\label{Q21} \\
	S(r)&=&\frac{1+\ell}{1-\frac{2 M}{r}+\frac{ 2(1+\ell)Q_0^2}{(2+\ell) r^2}-\frac{1}{3} (1+\ell)\Lambda _e r^2 },
	\label{Q22}\\
		\phi(r)&=&\frac{Q_{0}}{r}.
\end{eqnarray}

For convenience, we set $\Lambda_e=\frac{\kappa\lambda}{\xi}$ as an effective cosmological constant, which satisfies the relation $R_{\mu\nu}=\Lambda_e g_{\mu\nu}$ at the boundary of the spacetime and  is independent of the Lorentz-violating parameter.

 \begin{figure}
	\begin{center}
		\includegraphics[width=4cm]{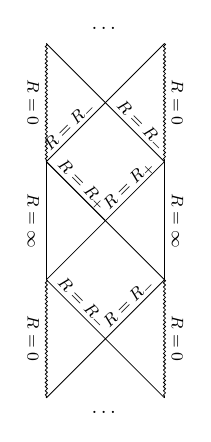}
	\end{center}
	\caption{ Penrose diagram of the non-extremal charged AdS black hole in bumblebee gravity.}
	\label{Penrose diagram2}
\end{figure}
As $\ell$ approaches to 0, this solution becomes RN-(A)dS black hole solution. Similarly, we compute the Kretschmann scalar in order to study the singularity. The expression for $K$ is 
\begin{eqnarray}
	K&=&\frac{8}{3} \Lambda _e^{2}+\frac{l}{(l+1) r^2}\Lambda _e-\frac{16 Q_{e}^2 \left((\ell+2) r (\ell r+12 M)-14 (\ell+1) Q_{e}^2\right)}{(\ell+1) (\ell+2)^2 r^8}\nonumber \\
	&+&\frac{4 \left(\ell^2 r^2+4 \ell M r+12 M^2\right)}{(\ell+1)^2 r^6},
\end{eqnarray}
which shows that, the RN-(A)dS-like singularity is only present at $r=0$ (assuming 
$0<\left|\ell\right|\ll 1$).

Through the coordinate transformation mentioned of $r$ ($r=\sqrt{\frac{1}{1+\ell}}R$), we obain

\begin{eqnarray}
	{d}{s}^{2}&=&-f(R){dt}^{2}+\frac{1}{f(R)}{dR}^{2}+\frac{1}{(1+\ell)}R^{2} {~d}\Omega ^{2},
	\label{pd1}
\end{eqnarray}
where $f(R)=(1-\frac{2\sqrt{1+\ell}M}{R}+\frac{ 2(1+\ell)^2 Q_0^2}{(2+\ell) R^2}-\frac{1}{3}\Lambda _eR^2 )$
then we can get the Penrose diagram of the non-extremal charged AdS black hole, as shown in Fig.~\ref{Penrose diagram2}, which is is identical to that of the classical RN-AdS spacetime. But the radii of the two-dimensional spheres at the same points on the Penrose diagrams are not the same.

\section{ SlOWLY ROTATING  CHARGED BLACK HOLE SOLUTIONS }
In this section, we will give slowly rotaing charged solutions through solving Einstein-bumblebee equations in both the cases with and without a cosmological constant. The forms of these solutions are very similar to  the Kerr-Newman (KN) and Kerr-Newman (KN)-(A)dS solutions.

In 1963, Kerr discovered the rotating black hole solution~\cite{Kerr:1963ud}. A few years later, Newman obtained  rotating charged black hole solution~\cite{Newman:1965my}.
In bumblebee gravity, some rotating solutions were also provided~\cite{Ding:2019mal,Ding:2020kfr,Jha:2020pvk}.
Here, we consider slowly rotating charged black hole solution with and without a cosmological constant. The general form of the corresponding metric is given by 
\begin{equation}
	ds^2=-A(r,\theta)dt^2+S(r,\theta)dr^2+2F(r)H(\theta)adtd\phi+\rho(r,\theta)^2d\theta^2+h(r,\theta)^2\sin^2\theta d\phi^2.
\end{equation}
The bumblee field and the $U(1)$ vector field are assumed as

\begin{eqnarray}
b_{\mu}&=&(0,b_r(r,\theta),0,0),\\
	F_{\mu\nu}&=&\partial_{\mu}A_\nu-\partial_{\nu}A_{\mu},\nonumber\\
A_{\mu}&=&(A_0(r,\theta),0,0,A_\phi(r,\theta)) ,
	\label{Fa}
\end{eqnarray}
\subsection{Case A: $V(X)=\frac{\lambda}{2}X^2$ and $ \Lambda=0$}

It is noted that when the Lorentz-violating parameter is 0, this solution should reduce to the Kerr-Newman solution. Similarly, when $a=0$, we should obtain the spherically symmetric black hole solution derived in the previous section. Therefore, we consider the product of $a$ and $\ell$ is a small quantity.

Ding et al. obtained a slowly rotating  black hole solution within bumblebee gravity~\cite{Ding:2019mal,Ding:2020kfr}. Following their approach and combining with Newman-Janis algorithm~\cite{Newman:1965tw,Newman:1965my,Janis:1965tx}, we obtained a slowly rotating charged black hole solution within bumblebee gravity. The metric in Boyer-Lindquist coordinates reads:
\begin{eqnarray}
	ds^2&=&-\frac{\Delta_r}{\rho^2}\Big(\mathrm{d}t-a\sqrt{1+\ell}\sin^2\theta\mathrm{d}\phi\Big)^2+(1+\ell)\frac{\rho^2}{\Delta_r}\mathrm{d}r^2+{\rho^2}\mathrm{d}\theta^2\nonumber\\
	&&\quad+\frac{\sin^2\theta}{\rho^2}\Big(a\sqrt{1+\ell}\mathrm{d}t-(r^2+a^2(1+\ell))\mathrm{d}\phi\Big)^2,\\
	A_0(r,\theta)&=&-\frac{Q_0r}{\rho^2},\\
	A_\phi(r,\theta)&=&\frac{Q_0a \sqrt{1+\ell}r \sin^2 \theta }{\rho ^2},\\
	b_r(r)&=&b\rho  \sqrt{\frac{ \left(1+\ell\right)}{\Delta _r}},
\end{eqnarray}where
\begin{eqnarray}
	\Delta _r&=&(1+\ell)a^2+r^2 -2 M r+\frac{2 (1 + \ell)Q_{0}^2}{(2+\ell)}, \nonumber\\
	\rho &=&\sqrt{r^2+(1+\ell)a^2 \text{cos}^2 \theta }	.
\end{eqnarray}
When $\ell\rightarrow0$, it recovers to the usual Kerr-Newman metric and when $a\rightarrow0$, it becomes to metric \eqref{qdc} with Eqs.~\eqref{Q11} and \eqref{Q12}. By solving $\Delta_{r}=0$, we can obatin the radii of the black hole's  horizon
\begin{equation}
	r_{\pm}=M \pm \sqrt{M^2-\frac{ (1+\ell) (2Q_0^2+(2+\ell)a^2)}{2+\ell}} .
\end{equation}
It becomes evident that the mass and spin parameters of the black hole must satisfy the following condition

\begin{equation}
	\frac{ (1+\ell) (2Q_0^2+(2+\ell)a^2)}{2+\ell}\leq M^2.
\end{equation}

Next, we discuss the small perturbations introduced by this rotating solution to the field equations, and impose certain constraints on the parameters through these perturbations.
We set

\begin{eqnarray}
	\Delta_{\mu\nu}=&R_{\mu\nu}&-\Lambda g_{\mu\nu}- \kappa\left[V'\left( 2 B_{\mu}B_{\nu}+b^{2}g_{\mu\nu}\right) +B_{\mu}^{\ \alpha}B_{\nu\alpha}+V g_{\mu\nu}\frac{1}{4}B_{\alpha\beta}B^{\alpha\beta}g_{\mu\nu} \right] \nonumber\\
	&-&\kappa(T^M_{\mu\nu}-\frac{1}{2}g_{\mu\nu}T^M)-
	\xi\left[\frac{1}{2}B^{\alpha}B^{\beta}R_{\alpha\beta}g_{\mu\nu}-B_{\mu}B^{\alpha}R_{\alpha\nu}-B_{\nu}B^{\alpha}R_{\alpha\mu}\right.\nonumber\\
	&+&\frac{1}{2}\nabla_{\alpha}\nabla_{\mu}\left(B^{\alpha}B_{\nu}\right)+\frac{1}{2}\nabla_{\alpha}\nabla_{\nu}\left(B^{\alpha}B_{\mu}\right)\left.-\frac{1}{2}\nabla^{2}\left(B_{\mu}B_{\nu}\right)\right] ,
	\label{TTb}      
\end{eqnarray}
naturally, for finite $Q_0$, when the field equations hold, we have $\Delta_{\mu\nu}=0$. Substituting the metric into the right hand side of  Eq.~\eqref{TTb}, our calculations yield:
\begin{equation}
	\Delta_{\mu\nu}=\left\{
	\begin{aligned}
\mathcal{O}(a^{2}\ell)&,&\quad  \mu\nu=tt,rr,\phi\phi,\theta\theta,r\theta,t\phi\\
		0&,&\quad  \text{otherwise}
	\end{aligned}	\label{Dr}
	\right.
	,
\end{equation}
At the same time, we derive the equations of motion for the bumblebee field and the electromagnetic field. Only the $r$ component  nonvanishing:

\begin{eqnarray}
	\left(\nabla_{\mu}B^{\mu r}-2( V'B^{r}-\frac{\xi}{2\kappa}B_\mu R^{\mu r}-\gamma B^{r}F^{\alpha\beta}F_{\alpha\beta} )\right)b&=&\mathcal{O}(a^{2}\ell),\nonumber\\
	\nabla_{\mu}\left(F^{\mu r}+\gamma B^{\alpha}B_{\alpha}F^{\mu r}\right)&=&\mathcal{O}(a^{2}\ell).
	\label{Dy}
\end{eqnarray}
Combining the above equations, we find that the rotating solution needs to satisfy the condition $\mathcal{O}(a^{2}\ell)$ being a small quantity. This implies that when considering this solution, $a$ and $\ell$ cannot both take large values simultaneously. Thus, for our solution, we need the condition that $a^2 \ell$ is small quantities ($\frac{a^2  \left|\ell\right|}{M^2}\ll 1$) for finite $Q_0$. 

Theoretically, there is no constraint on the spin parameter $ a $ and the Lorentz-violating parameter $ \ell $. The constraint $a^2\left|\ell\right| \ll M^2$ or $a^2\left|\ell\right| \ll 1$ for $M=1$ on the  parameters  is imposed to ensure that our analytical solution approximately satisfies the field equations. We will use the analytical solution to study some observables of the black hole shadow, under the condition $a^2\left|\ell\right| \ll 1$. This constraint affects the range of the parameter combination $ a^2 \ell $, but it does not affect the individual parameter values.  As $ \ell $ becomes  sufficiently small, $ a $ may take larger values and even approach $M$. Conversely, when $ a $ is sufficiently small, $ \ell $ may take larger values. However, based on previous studies, it has been found that $ \ell $ is constrained to a small value~\cite{Casana:2017jkc}, so in astrophysical scenarios, we consider $ \ell $ to be sufficiently small, allowing a large range for $ a $, which can even approach $M$.

 When $ \ell $ is sufficiently small, we can consider the extreme black hole case. The extreme black hole condition is approximately given by $\frac{(1+\ell) (2Q_0^2 + (2+\ell) a^2)}{2+\ell} = M^2$, since our solutions are not exact. For sufficiently small $ \ell $, the extreme black hole condition can be approximated as $Q_0^2 + a^2 = M^2.$ From this, we can see that $ a^2 \leq M^2 $. As long as $ \ell $ is sufficiently small, $ a^2 \left|\ell\right| \ll M^2 $ is still satisfied. In summary, provided that $ \ell $ remains sufficiently small, our solution remains valid, even for the near-extremal KN-like black hole case.

\subsection{Case B: $V(X)=\frac{\lambda}{2}X$ and $ \Lambda\neq0$}
Next, we consider the case of $V(X)=\frac{\lambda}{2}X$ and $ \Lambda\neq0$. Similar to the case A, we can get the following solution:
\begin{eqnarray}
	ds^2&=&-\frac{\Delta_r}{\rho^2}\Big(\mathrm{d}t-\frac{a\sqrt{1+\ell}\sin^2\theta}{\Xi}\mathrm{d}\phi\Big)^2+(1+\ell)\frac{\rho^2}{\Delta_r}\mathrm{d}r^2+\frac{\rho^2}{\Delta_\theta}\mathrm{d}\theta^2\nonumber\\
	&&\quad+\frac{\sin^2\theta\Delta_\theta}{\rho^2}\Big(a\sqrt{1+\ell}\mathrm{d}t-\frac{r^2+a^2(1+\ell)}{\Xi}\mathrm{d}\phi\Big)^2,\\
A_0(r,\theta)&=&-\frac{Q_0r}{\rho^2},\\
A_\phi(r,\theta)&=&\frac{Q_0ar \sin^2 \theta }{\Xi  \rho ^2},\\
b_r(r)&=&b\rho  \sqrt{\frac{ \left(1+\ell\right)}{\Delta _r}},
\end{eqnarray}
where
\begin{eqnarray}
	\Delta _{\theta }&=&1+\frac{1}{3}(1+\ell)^{2} \Lambda_{e}a^2 \text{cos}^2 \theta , \nonumber\\
	\Delta _r&=&\left((1+\ell)a^2+r^2\right) \left(1-\frac{1}{3}(1+\ell) \Lambda_{e}r^2\right)-2 M r+\frac{2 (1 + \ell)Q_0}{(2+\ell)} ,\nonumber\\
	\Xi &=&1+\frac{1}{3}(1+\ell)^{2} \Lambda_{e}a^2 ,\nonumber\\
	\rho &=&\sqrt{r^2+(1+\ell)a^2 \text{cos}^2 \theta }	.
\end{eqnarray}
From the form of the solution, it is a KN-(A)dS-like solution. Note that when the charge parameter  $Q_\text{0}=0$, the solution is reduced to a slowly rotating Kerr-(A)dS-like black hole solution, which is not found in previous work. Otherwise, when $\ell=0$ and $a=0$, we get the usual KN-(A)dS black hole solution and the metric  \eqref{qdc} with \eqref{Q21} and \eqref{Q22}, respectively. 

By calculating the errors in the field equations, we find that the results are similar to case A. 
Similarly, the rotating solution must satisfy the condition that $a^{2}\ell$ is a small quantity.

\section{  SHADOW OBSERVABLES }

The black hole shadow is formed by the strong gravitational field of the black hole. The shadow appears as a two-dimensional dark region to a distant observer. Shadows serve as one of the most intuitive pieces of evidence for the presence of black holes. Particularly, the unveiling of an image provides even stronger evidence for black hole's  existence~\cite{EventHorizonTelescope:2019dse,EventHorizonTelescope:2019ggy,EventHorizonTelescope:2019pgp,EventHorizonTelescope:2019ths,EventHorizonTelescope:2022wkp}. 
The shadow of the Kerr-like black hole in bumblebee gravity has been extensively studied in previous works~\cite{Wang:2021irh,Ding:2019mal,Kuang:2022xjp,Chen:2020qyp,Islam:2024sph,Afrin:2024khy}.
In this section, we aim to investigate the shadow of the KN-AdS like black holes in the framework of Einstein-bumblebee gravity.
\subsection{The photon orbit}
In this section, we will derive the equations of motion of photons orbiting the black holes. From Eqs.~\eqref{Dr} and~\eqref{Dy}, it is apparent that the field equations hold when the rotating black hole solution satisfies that 
$a\ell^2$ is
a small quantity. Therefore, in the subsequent part, all the analysies are under this condition. The motion of a photon in the background of the black hole is given by the geodesics. However, the process of solving geodesic equations is very difficult, we choose to use the Hamilton-Jacobi approach to solve it.
The Lagrangian for the photon is
\begin{equation}
	\mathcal{L}=\frac{1}{2}g_{\mu\nu}\dot{x}^{\mu}\dot{x}^{\nu},
\end{equation}
where $\dot{x}$ is defined as $\dot{x}^{\mu}=dx^{\mu}/d\lambda=u^{\mu}$, with the four-velocity of the photon $u^{\mu}$ and  the affine parameter  $\lambda$. To facilitate the study of photon orbits and the black hole shadow observed by a distant observer, we introduce two parameters:
\begin{equation} \label{bardeen-00}
	\xi = \frac{L_{z}}{E} \, , \quad \eta = \frac{K}{E^2}  \, ,
\end{equation}
where $E$, $L_{z}$, and $K$ represent the energy, the axial component of the angular momentum, and the Carter constant~\cite{Carter:1968rr}, respectively. The expressions of the energy and  the $z$-component of the angular momentum is given by
\begin{equation}
	E=-\frac{\partial{\mathcal{L}}}{\partial\dot{t}}=-g_{\varphi t}\dot{\varphi}-g_{tt}\dot{t},
\end{equation}

\begin{equation} 
	L_z=\frac{\partial{\mathcal{L}}}{\partial\dot{\varphi}}=g_{\varphi\varphi}\dot{\varphi}+g_{\varphi t}\dot{t}.
\end{equation}
Using the Hamilton-Jacobi equation, we can get the null geodesic equation of the photon around the slowly rotating black hole:
\begin{eqnarray}
	\dot{t}&=&\frac{E}{\rho^2}\Bigg[\frac{\left(r^2+(1+\ell)a^2\right)\left[\left(r^2+(1+\ell)a^2\right)-\sqrt{1+\ell}a\xi\Xi\right]}{\Delta_r} \\
	&&+\frac{\sqrt{1+\ell}a\left(\xi\Xi-\sqrt{1+\ell}a\sin^2\theta\right)}{\Delta_\theta}\Bigg], \nonumber \\
	\dot{\varphi}&=&\frac{E\Xi}{\rho^2}\left[\frac{\sqrt{1+\ell}a(\left(r^2+(1+\ell)a^2\right)-\sqrt{1+\ell}a\xi\Xi)}{\Delta_r}+\frac{\xi\Xi-\sqrt{1+\ell}a\sin^2\theta}{\sin^2\theta\Delta_\theta}\right],\\
	(1+\ell)(\rho^2\dot{r})^2&=&E^{2}\left[\left[\left(r^{2}+(1+\ell)a^{2}\right)-\sqrt{1+\ell}a\xi\Xi\right]^{2}-\Delta_{r}\Big[\frac{\left(\sqrt{1+\ell}a-\xi\Xi\right)^{2}}{\Delta_{\theta}}+\eta\Big]\right]\nonumber \\
	&\equiv&(1+\ell)\mathcal{R}(r), \\
	\rho^2\dot{\theta}&=&E^2\left[\eta\Delta_{\theta}-\cos^2\theta\left(\frac{\xi^2\Xi^2}{\sin^2\theta}-(1+\ell)a^2\right)\right]\equiv\Theta(\theta).
	\label{tpr}
\end{eqnarray}

\begin{figure}
	\begin{center}
		\includegraphics[width=7cm]{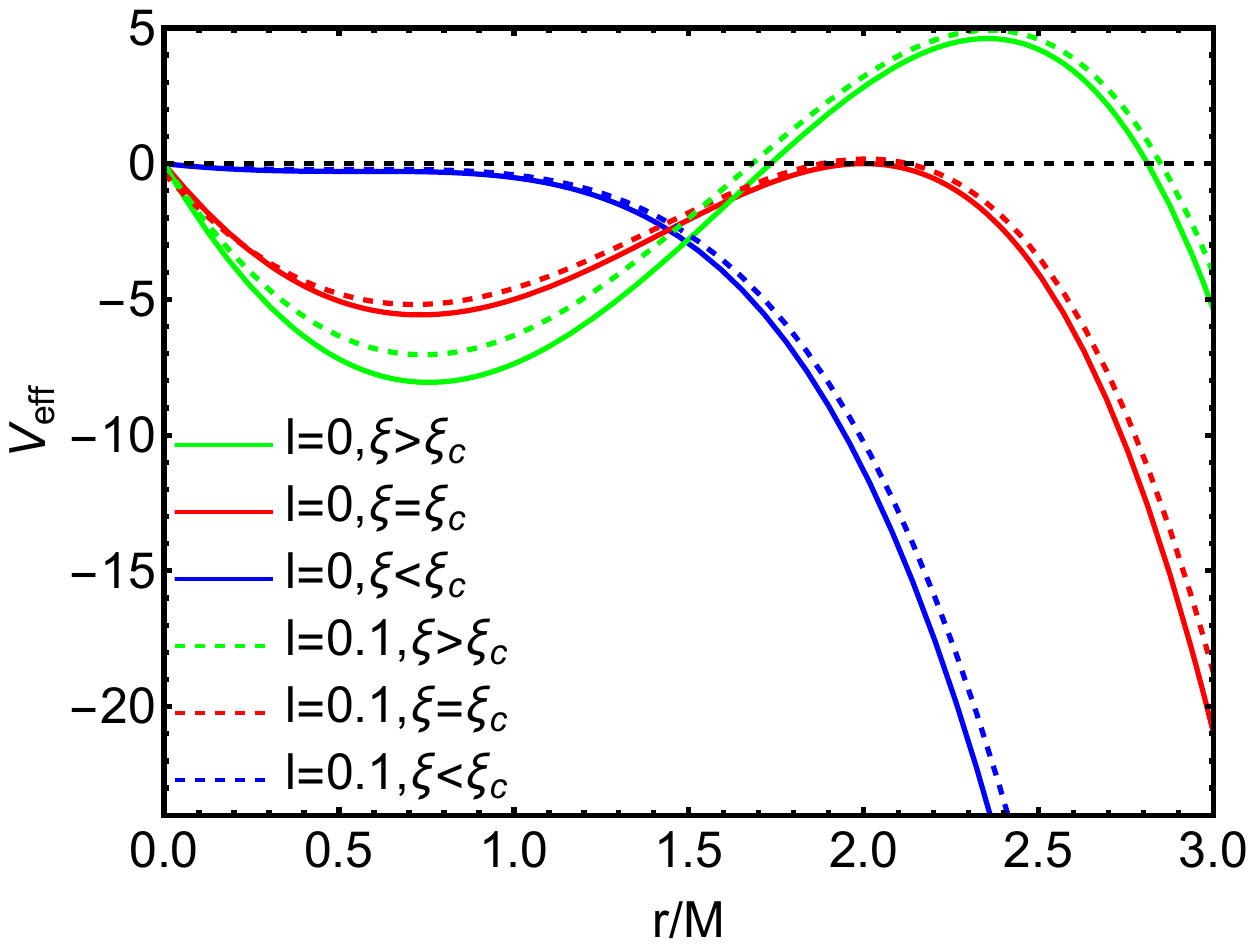}
	\end{center}
	\caption{ The effective potential $V_\text{eff}$ for the photon. The paramengts are set as $M=1,a = 0.1$ , $Q_0=0.1$ , $l=0,0.1$ and  $\xi_c$ is an the value of $\xi$ for unstable circular null orbit.}
	\label{Veff}
\end{figure}
 Photons with sufficiently large orbital angular momentum can escape from the black hole, while those with small angular momentum will fall into the black hole. Now we focus on the critical case, the circular null orbit, where the photons will neither escape from nor fall into the black hole. The radial motion can also be expressed as
\begin{equation}
(\rho^2\dot{r})^2+V_\text{eff}=0,
\end{equation}
where the effective potential $V_\text{eff}$ reads
\begin{equation}
	V_\text{eff}=-\frac{E^{2}}{(1+\ell)}\left[\left[\left(r^{2}+(1+\ell)a^{2}\right)-\sqrt{1+\ell}a\xi\Xi\right]^{2}-\Delta_{r}\big[\frac{\left(\sqrt{1+\ell}a-\xi\Xi\right)^{2}}{\Delta_{\theta}}+\eta\big]\right].
\end{equation}
In Fig.~\ref{Veff}, we depict the effective potential $V_\text{eff}$ for a rotating charged  black hole 
with $\ell=0$ and $\ell=0.1$. 
From Fig.~\ref{Veff}, we can observe that Lorentz-violating parameter greater than 0 can increase the effective potential~\cite{Wei:2015dua}.
For the unstable circular null orbit, the effective potential satisfies
\begin{equation} 
	{V}(r)=\frac{d{V}(r)}{dr}=0, ~\text{and} ~\frac{d^2 V(r)}{dr^2}<0,
\end{equation}
 then we can obtain~\cite{Tsukamoto:2014tja,Tsukamoto:2017fxq}

\begin{eqnarray}
	\eta&=&\frac{r^2\left(16(1+\ell)a^2\Delta_r\Delta_\theta-\left(r\Delta_r'-4\Delta_r\right)^2\right)}{(1+\ell)a^2\Delta_\theta\Delta_r'^2},\label{eta}\\
	\xi&=&\frac{\left(r^2+(1+\ell)a^2\right)\Delta_r'-4r\Delta_r}{\sqrt {1+\ell}a\Xi\Delta_r'}.
	\label{xi}
\end{eqnarray}

\begin{figure}
	\begin{center}
		\subfigure[\ $M=1,a=0.1,\ell=0.01,Q_{0} =0.2,\theta =\frac{\pi }{2}.$
		\label{cp22}]{\includegraphics[width=5.5cm]{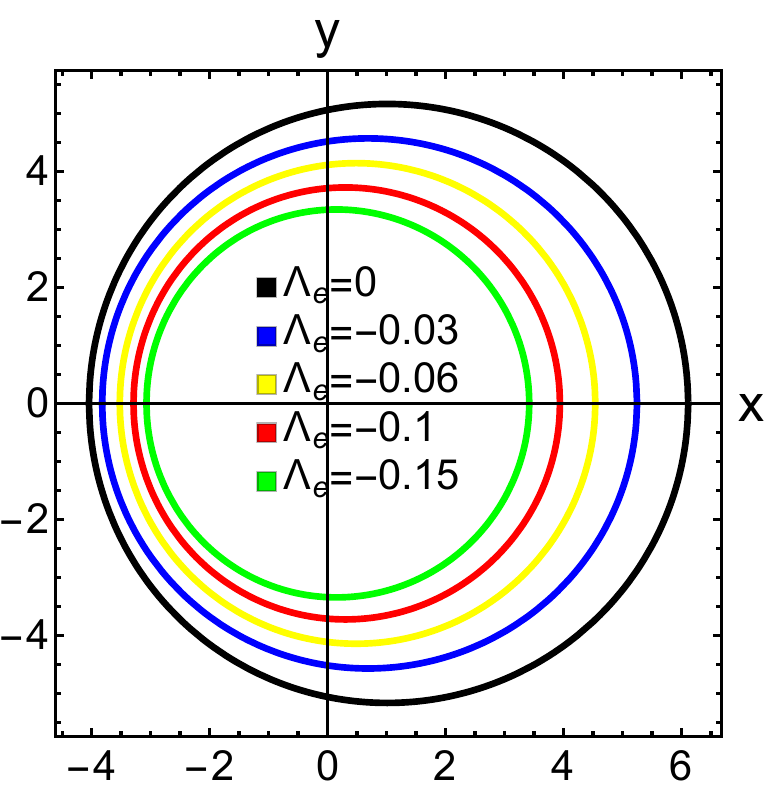}}
		\  \
		\subfigure[\ $M=1,a=0.9,\ell=0.01,\Lambda _{e}=-0.1,Q_{0} =0.3.$ \label{cp23}]{\includegraphics[width=5.5cm]{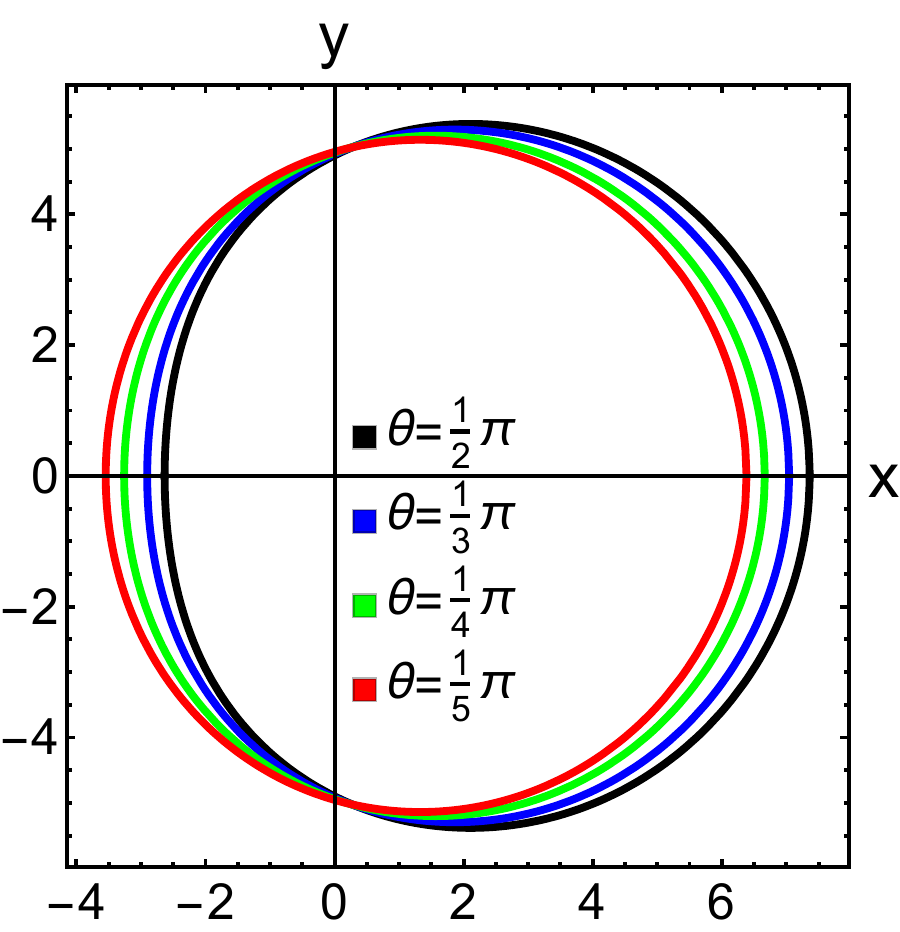}}
		\ \
	\end{center}
	\begin{center}
		\subfigure[\ $M=1,a=0.1,\Lambda _{e}=-0.1,Q_{0} =0.2.$
		\label{cp24}]{\includegraphics[width=5.5cm]{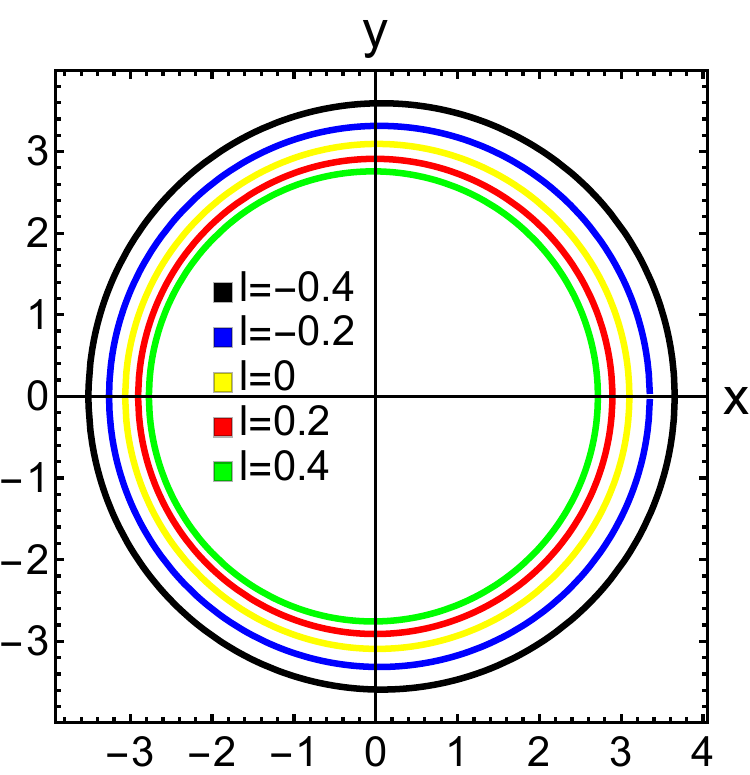}}
		\  \
		\subfigure[ \ $M=1,a=0.647,\ell=0.01,\Lambda _{e}=-0.2,\theta =\frac{\pi }{2}.$ \label{cp25}]{\includegraphics[width=5.5cm]{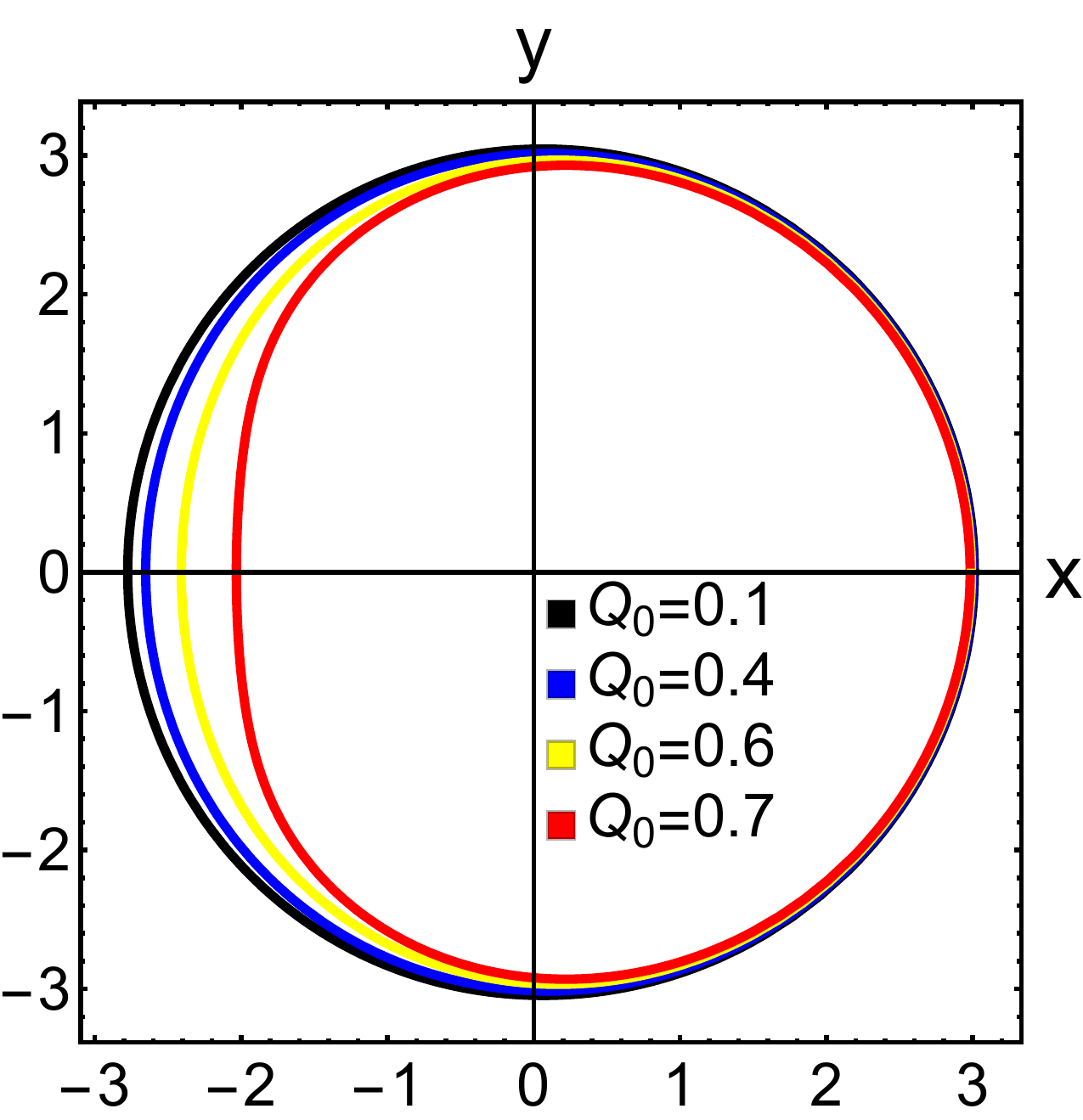}}

	\end{center}
	\caption{The shadows for the  KN-(A)dS-like black hole with different parameters.}
	\label{shadow}
\end{figure}

\subsection{Observables}
We now aim to determine the apparent shape of the slowly rotating charged black hole's shadow.   The apparent shape of the black hole shadow for an observer far away from the black hole can be described using the celestial coordinates $x$ and $y$:

\begin{eqnarray}
	x&=&\lim_{r\to\infty}\frac{-rp^{(\varphi)}}{p^{(t)}}=-\xi\csc\theta,\\
	y&=&\lim_{r\to\infty}\frac{rp^{(\theta)}}{p^{(t)}}=\pm \sqrt{\eta+a^2\cos^2\theta-\xi^2\cot^2\theta}.	
\end{eqnarray}
The coordinate $x$ denotes the apparent perpendicular distance from the shape's projection on the equatorial plane, while $y$ represents the apparent perpendicular distance of the shape from the axis of symmetry.
The boundary curve of the shadow is determined by the  unstable circular null orbit. Substituting the metric into Eqs.~\eqref{eta} and~\eqref{xi}, we have
\begin{eqnarray}
	\eta&=&\frac{4r^2 \left( \left(a^2 (\ell+1)+r^2\right) \left(1-\frac{1}{3} (\ell+1) r^2 \Lambda _e\right)-2Mr+\frac{(1+\ell)Q_{0}^2}{(1+\frac{\ell}{2})}\right)}{ \left(\frac{1}{3} (\ell+1) r \Lambda _e \left(a^2 (\ell+1)+2 r^2\right)+M-r\right)^2} \nonumber\\
	&&-\frac{ r^2 \left(-\frac{1}{3} a^2 (\ell+1) r^2 \Lambda _e+2 a^2 -\frac{3 M r}{1+\ell}+\frac{4Q_{0}^2}{(2+\ell)}+r^2\right)^2}{ a^2  \left(\frac{1}{3} a^2 (\ell+1)^2 \Lambda _e \cos ^2(\theta )+1\right) \left(\frac{1}{3} (\ell+1) r \Lambda _e \left(a^2 (\ell+1)+2 r^2\right)+M-r\right){}^2},\\
	\xi&=&\frac{\frac{1}{3} a^2 (\ell+1)^2 r \Lambda _e \left(a^2 (\ell+1)+r^2\right)+a^2 (\ell+1) (M+r)+r \left(r (r-3 M)+\frac{4Q_{0}^2}{(2+\ell)}\right)}{a \sqrt{\ell+1} \left(\frac{1}{3} a^2 (\ell+1)^2 \Lambda _e+1\right) \left(\frac{1}{3} (\ell+1) r \Lambda _e \left(a^2 (\ell+1)+2 r^2\right)+M-r\right)}.
\end{eqnarray}

\begin{figure}
	\begin{center}
		\subfigure[ \ $\theta =\frac{\pi }{2}$ \label{Q}]{\includegraphics[width=5cm]{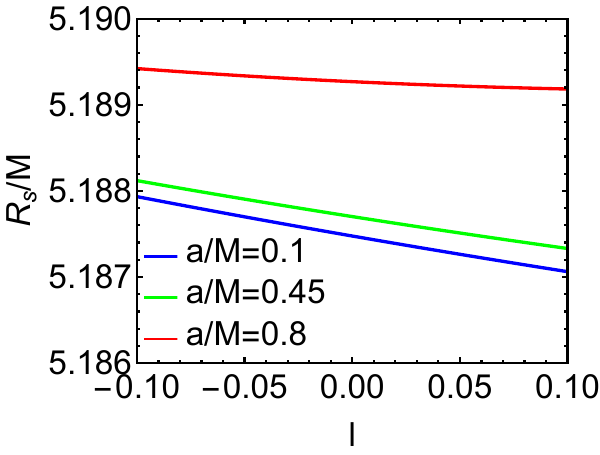}}
		\  \
		\subfigure[\ $\theta =\frac{\pi }{4}$ \label{cp21}]{\includegraphics[width=5cm]{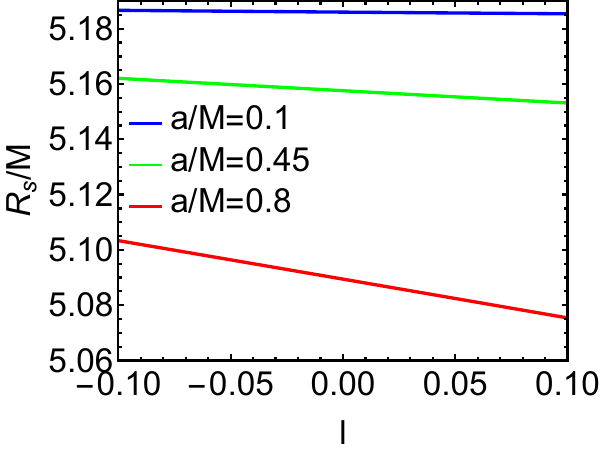}}
		\ \
		\subfigure[\ $\theta =\frac{\pi }{6}$
		\label{cp12}]{\includegraphics[width=5cm]{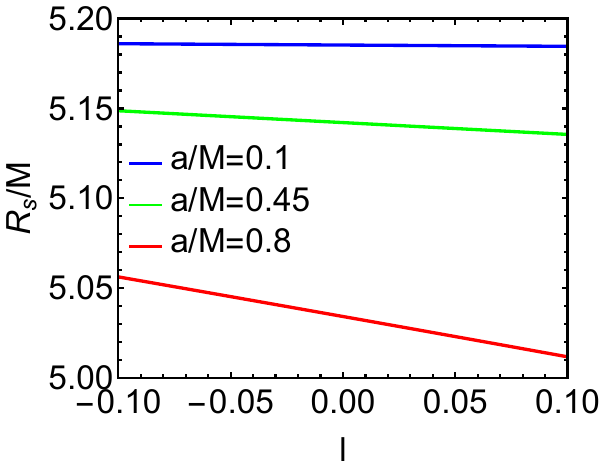}}
		\  \
		\caption{ The radius $R_s$ of the shadow against the Lorentz-violating parameter $\ell$.}
		\label{p3}	
	\end{center}

	\begin{center}
		\subfigure[ \ $\theta =\frac{\pi }{2}$ \label{Q}]{\includegraphics[width=5cm]{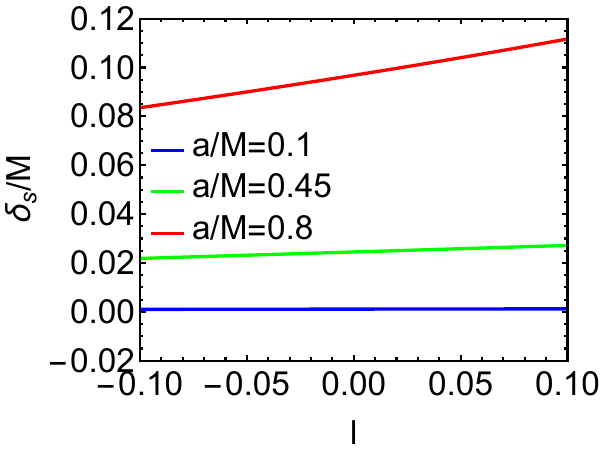}}
		\  \
		\subfigure[\ $\theta =\frac{\pi }{4}$ \label{cp21}]{\includegraphics[width=5cm]{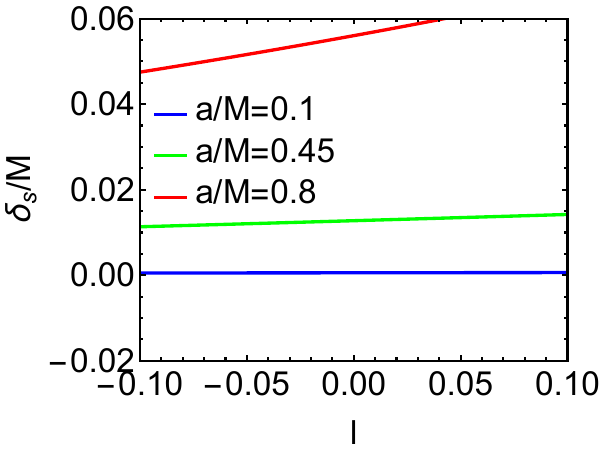}}
		\ \
		\subfigure[\ $\theta =\frac{\pi }{6}$
		\label{cp12}]{\includegraphics[width=5cm]{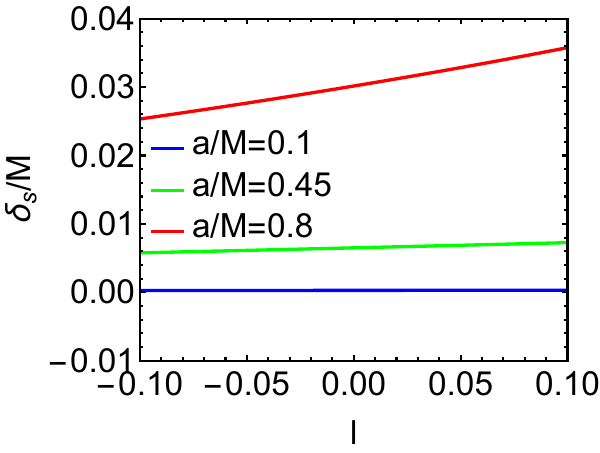}}
		\  \
		
	\end{center}
	\caption{The distortion parameter $\delta_s$ against the Lorentz-violating parameter $\ell$.}
	\label{p4}	
	\end{figure}

It is easy to see  that both the coordinates $x$ and $y$ are the
functions of $(M,Q_{eff},a,\Lambda_e,\ell,\theta)$. We set $M=1$, and show the shapes of the shadow in Fig.~\ref{shadow}. Similar to the KN-AdS black hole, the shadow radius increases with the negative effective cosmological costant $\Lambda_e$, as shown in Fig.~\ref{cp22}.  From Fig.~\ref{cp23} we can see that, the position of the shadow depends on the observation angle $\theta$. The Lorentz-violating parameter $\ell$ also affects the shaow size which can be seen from Fig.~\ref{cp24}. Besides, the deformation of the shadow becomes more pronounced when the charge parameter $Q_0$ increases, which can be seen from Fig.~\ref{cp25}.

\begin{figure}
	\begin{center}
		\subfigure[ \  $\theta =\frac{\pi }{2}$ \label{Q}]{\includegraphics[width=5cm]{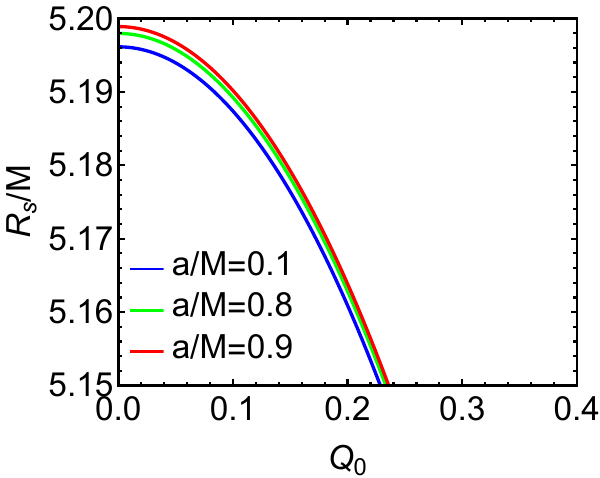}}
		\  \
		\subfigure[\  $\theta =\frac{\pi }{4}$ \label{cp21}]{\includegraphics[width=5cm]{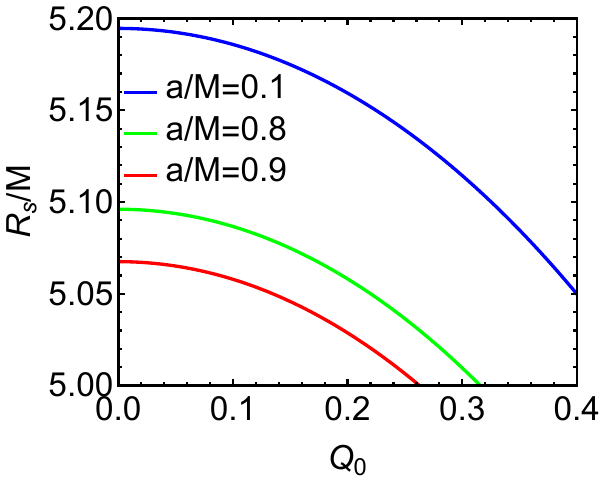}}
		\ \
		\subfigure[\ $\theta =\frac{\pi }{6}$ 
		\label{cp12}]{\includegraphics[width=5cm]{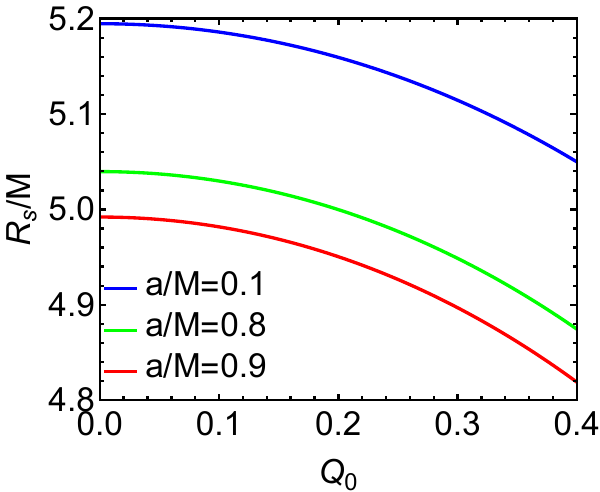}}
		\  \
		
	\end{center}
	\caption{The radius $R_s$ of the shadow against the charge parameter $Q_0$.}
	\label{p5}

	\begin{center}
		\subfigure[ \ $\theta =\frac{\pi }{2}$  \label{Q}]{\includegraphics[width=5cm]{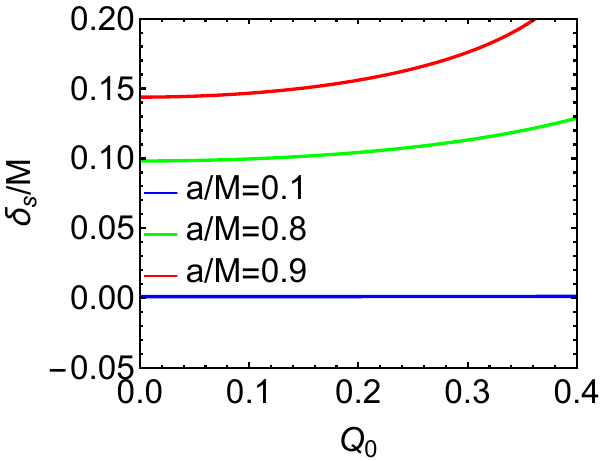}}
		\  \
		\subfigure[\ $\theta =\frac{\pi }{4}$ \label{cp21}]{\includegraphics[width=5cm]{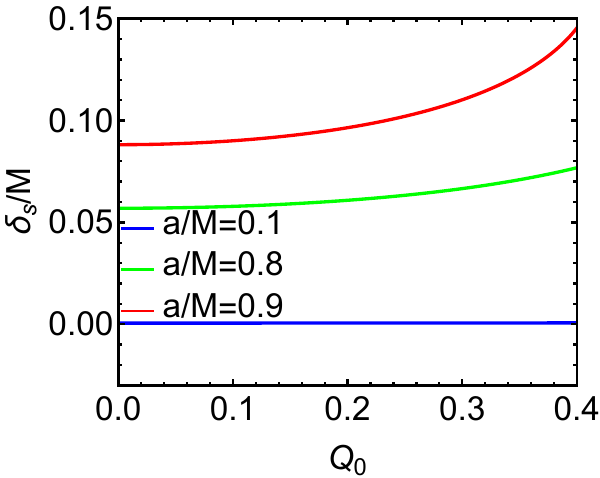}}
		\ \
		\subfigure[\ $\theta =\frac{\pi }{6}$ \label{cp12}]{\includegraphics[width=5cm]{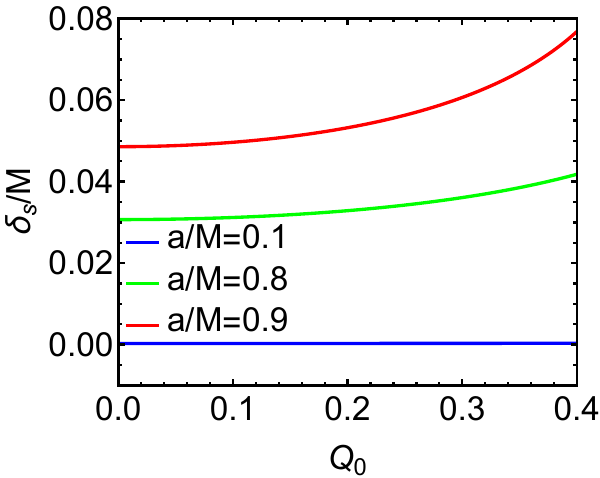}}
		\  \
	
	\end{center}
	\caption{ The distortion parameter $\delta_s$ against the charge parameter $Q_0$.}
	\label{p6}	
\end{figure}

To effectively analyze astronomical observational data, it is essential to establish easily measurable and reliable astronomical observables. In this study, we will examine the following observables, the radius $R_s$ and the distortion parameter $\delta_s$ which are defined by Hioki and Maeda~\cite{Hioki:2009na}. The parameter $R_s$
represents the radius of the reference circle, which is expected to pass through the top point $(x_t,y_t)$, the bottom point $(x_b,y_b)$, and the right point $(x_r,0)$ of the shadow. Meanwhile, the distortion parameter 
quantifies the distortion of the black hole shadow in comparison with the reference circle, and they can be expressed as
\begin{equation}
	R_s=\frac{(x_t-x_r)^2+y_t^2}{2(x_r-x_t)},\quad\delta_s=\frac{x_l-\tilde{x}_l}{R_s},
\end{equation}
where $(x_l,0)$ and $(\tilde{x}_l,0)$ represent the left points of the shadow and the reference circle, respectively.

For a given spin parameter, we illustrate the radius $R_s$ and the distortion parameter  $\delta_s$ as functions of the Lorentz-violating parameter $\ell$ for $\theta=\frac{\pi}{2},\theta=\frac{\pi}{4},\theta=\frac{\pi}{6}$.
The behavior of $R_s$ and $\delta_s$ is exhibited in Fig.~\ref{p3} and Fig.~\ref{p4}. For  fixed  $a$ and $\theta$, the radius $R_s$ decreases with $\ell$, but $\delta_s$ increases with $\ell$ increases. Particularly noteworthy is the observation that for observers situated far from the equatorial plane, the shadow of a rapidly rotating black hole appears smaller compared to that of a slower one with the same $\ell$. However, this discrepancy diminishes as the observation angle increases, eventually resulting in a convergence of all curves as $\theta$ approaches $\frac{\pi}{2}$. Furthermore, as the observer progressively approaches the equatorial plane, the distortion of the shadow becomes increasingly pronounced.

Similarly, we investigate the relationship between the value of the charge parameter and  $R_s$ and $\delta_s$ which is shown in Fig.~\ref{p5} and Fig.~\ref{p6}. We find that $R_s$ decreases with the charge parameter, whereas $\delta_s$ increases with the charge parameter. Another notable observation in our study is that for small values of the spin parameter $a$, the deformation of the shadow is nearly imperceptible.

The shadow image of the supermassive black hole M87* was  photographed by the EHT collaboration with the mass $M=6.5\times10^9M_\odot$, the distance $r_0=16.8Mpc$, and the inclination angle $\theta_0=17^o$~\cite{EventHorizonTelescope:2019dse,EventHorizonTelescope:2019ggy,EventHorizonTelescope:2019pgp,EventHorizonTelescope:2019ths,EventHorizonTelescope:2022wkp}.  And in Ref.~\cite{EventHorizonTelescope:2021dqv}, the authors stated that the shadow size of M87* should lie between $4.31M$ and $6.08M$. We will use the shadow observable $R_a$ to constrain the parameters of the Kerr-Newman-like  black hole in bumblebee gravity. The definition of the characteristic areal radius of the shadow curve can be expressed as~\cite{Banerjee:2019nnj,Abdujabbarov:2015xqa}
\begin{eqnarray}
	R_a=\sqrt{\frac{2}{\pi}\int^{r_{max}}_{r_{min}}\Big(y(r)\frac{dx(r)}{dr}\Big)}. 
\end{eqnarray}
\begin{figure}
	\begin{center}
		\includegraphics[width=70mm]{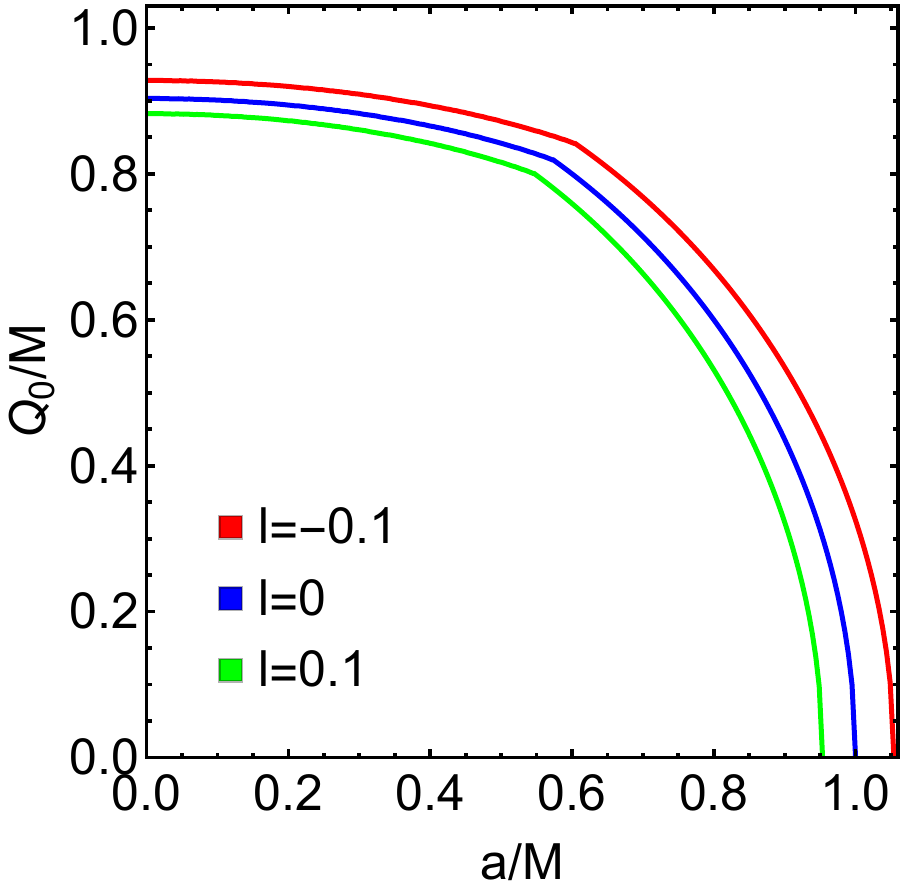}
	\end{center}
	\caption {The constraints of the EHT observation $4.31M \leq R_a\leq6.08M$ on the parameters ($a,Q_0$). The real curves represent $R_a=4.31M$ with different Lorentz-violating parameters $\ell$.} \label{xianzhiae}
		\label{p9}
\end{figure}Figure~\ref{p9}  shows the constraints on the black 
	hole parameters by the EHT observation with the shadow size $R_a$. All the solid lines represent the lower bound $ R_a = 4.31M $ or $\frac{(1+\ell) (2Q_0^2 + (2+\ell) a^2)}{2+\ell} = M^2$. The reasonable parameters, subject to the constraints, lie below these lines.
 The results illustrate that the larger the Lorentz-violating parameter $\ell$, the stronger the constraint on the rotating parameter $a$ and charge parameter $Q_0$. For  $\ell=-0.1$ and $\ell=0.1$, the ranges of the charge paramter $Q_0$ are $Q_0/M\in(0,0.93)$ and $Q_0/M\in(0,0.88)$, respectively.  When we  consider a small Lorentz-violating parameter, according to the range of the shadow size $R_{a}$, we can not make any constraint on the rotating parameter $a$.

\section{ CONCLUSIONS }
In this paper, by considering the coupling between the electromagnetic field and the bumblebee vector field, we obtained the RN-like black hole solutions and the RN-(A)dS-like black hole solutions in bumblebee gravity. When the Lorentz-violating parameter vanishes $(\ell = 0)$, these solutions reduce to the RN and RN-(A)dS black hole solutions. In the stationary and axially symmetric scenario, we  obtained solutions for slowly rotating charged black holes, i.e., the  KN-(A)dS-like black hole,  within the framework of bumblebee gravity. We further analyzed the series expansion of small quantities.

 Because the nonminimal interaction term between the electromagnetic field and bumblebee field was introduced, the  Maxwell's equations was also modified. The charge term of the black hole solution is related to the Lorentz-violating parameter. When the electromagnetic field does not exist, our solutions are then reduced to another bumblebee black hole solutions in previous works. 
 
Moreover, in the background of the  KN-(A)dS-like black hole, we derived null geodesics using the Hamilton-Jacobi equation. Utilizing the conditions for unstable circular null orbits, we derived two conserved parameters, $\xi$ and $\eta$, which govern the photon's motion. Using these parameters, we determined the celestial coordinates of the shadow. Subsequently, we investigated the size and deformation of the shadow. By varying the observation angle $\theta=\frac{\pi}{2},$ $\theta=\frac{\pi}{4},$ $\theta=\frac{\pi}{6}$, we calculated the shadow's radius $R_s$ for different parameters. We observed that for fixed spin parameter $a$ and the Lorentz-violating parameter $\ell$, observers situated at lower latitudes consistently observe a larger shadow. Meanwhile, we found that the radius of the reference circle decreases with the Lorentz-violating parameter and electric charge , while the distortion parameter increases with the Lorentz-violating parameter and electric charge.

\section*{  Acknowledgements }
	We are grateful to Tao-Tao Sui and Ke Yang for useful discussions. This work was supported by 
	the National Key Research and Development Program of China (Grant No. 2021YFC2203003), 
	the National Natural Science Foundation of China (Grants No. 12475056, No. 12247101, No. 12475055 and No. 12205129), 
	the 111 Project (Grant No. B20063), 
	Gansu Province's Top Leading Talent Support Plan.

 \appendix
 \section{ A BRIEF DISCUSSION ON BLACK HOLE THERMODYNAMICS }	In this  appendix, we  briefly discuss the impact of $\ell$ $(\ell=\xi b^2)$ on black hole thermodynamics.
 First, using the surface gravity, the black hole temperature is given by
 \begin{eqnarray}
	T_h=\frac{-2{Q_0}^2 (1+\ell)-{r_h}^2 (2+\ell) \left(\Lambda_e{r_h}^2 (1+\ell )-1\right)}{4 \pi {r_h}^3 \sqrt{1+\ell } (2+\ell )},
\end{eqnarray} 
	 where $ r_h $ is the event horizon radius. 
	Noting  that when  $\left| \ell \right| <1 $, we can deduce that by redefining certain physical quantities ($Q=\sqrt{\frac{2(1+\ell) }{2+\ell}}{Q_0}$, $\Lambda=(1+\ell )\Lambda_e$, $T_h'=\sqrt{1+\ell}T_h$), the black hole temperature can be restored to the original form of the RN-AdS black hole. This means that the curve of the temperature as a function of the horizon radius does not change fundamentally. For the RN-AdS-like black hole in bumblebee gravity, the extended first law of thermodynamics can be formulated as
\begin{equation}dM=T_h dS+\Phi_0 dQ_0+\frac{\Theta}{8\pi}d\Lambda_e,\end{equation}	where $S=4\pi\sqrt{1+\ell} r_h^2$, $\Theta=-\frac{4}{3}(1+\ell)\pi r_h^3$, $\Phi_0=\frac{2(1+\ell)Q_0 }{(2+\ell)r}$.
 	This is similar to the first law of thermodynamics for the RN-AdS black hole.

Furthermore, by using the critical point equations 
 	\begin{equation}\partial_{r_{h}}T\left(r_{h}\right)=0,\quad\partial_{r_{h}}^{2}T\left(r_{h}\right)=0,\end{equation}
 	 we can easily obtain $Q_0= \frac{\sqrt{2+\ell} r_h} {\sqrt{6(1+\ell) }}$ and $\Lambda_e =\frac{1}{2 {r_h}^2 (1+\ell )}$ which only differ from the critical values in the RN-AdS black hole by a coefficient. Therefore, the critical point has indeed been modified, but there is no essential difference from the RN-AdS black hole.

 The specific heat can be calculated as 
 		\begin{eqnarray}
 			C_\text{B}=T_h\left(\frac{\partial S}{\partial T_h}\right)_{Q_0,\Lambda_e}=\frac{2 \pi  \sqrt{1+\ell } r_h^2 \left(-\left(\Lambda_e  (1+\ell ) r_h^4\right)+r_h^2-\frac{2 {Q_0}^2 (1+\ell )}{\ell +2}\right)}{\frac{6 {Q_0}^2 (1+\ell )}{2+\ell}-r_h^2 \left(\Lambda_e  (1+\ell ) r_h^2+1\right)}. 
 	\end{eqnarray}It can be  seen that the specific heat has been modified. By  redefining certain physical quantities ($Q=\sqrt{\frac{2(1+\ell) }{2+\ell}}{Q_0}$, $\Lambda=(1+\ell )\Lambda_e$), the specific heat becomes equal to that of the RN-AdS black hole, multiplied by a factor, satisfying \( C_{\text{B}} =\sqrt{1+\ell } C_{\text{R}} \).  It is found  that for $1+12Y>0$, where $Y=\frac{(1+\ell )^2{Q_0}^2\Lambda_e }{(1+\frac{\ell}{2})}$, the heat capacity is positive within the intervals $r_{h}\subset\left(\sqrt{\frac{\sqrt{1-4Y}-1}{-2(1+\ell )\Lambda_e}},\sqrt{\frac{1-\sqrt{1+12Y}}{-2(1+\ell )\Lambda_e}}\right)\cup\left(\sqrt{\frac{1+\sqrt{1+12Y}}{-2(1+\ell )\Lambda_e}},\infty\right)$. 
 	 While $1+12Y\leq0$, the heat capacity is positive for	$r_{h}\subset\left(\sqrt{\frac{\sqrt{1-4Y}-1}{-2(1+\ell )\Lambda_e}},\infty\right)$. From this, we conclude that stability of black hole thermodynamics does not change fundamentally. On the other hand, from the black hole free energy 
 	\begin{eqnarray}
 		F=M-T_{h}S=\frac{1}{4} \left(\frac{3 (1+\ell) Q_0^2}{\left(1+\frac{\ell}{2}\right) r_h}+3(1+\ell)\Lambda_e r_h^3+r_h\right),
 	\end{eqnarray}it reveals the possible black hole phase transitions are similar to those of the RN-AdS black hole, the system may exhibit black hole phases with the following stability: small black holes (stable), medium black holes (unstable), and large black holes (stable).

From these aspects, we can conclude that the nonminimal coupling does indeed modify the black hole thermodynamics, but it does not change some of the essential behaviors. Additionally, the modification parameter is generally small, indicating that Lorentz symmetry breaking has a relatively minor impact on black hole thermodynamics in this model.
\bibliography{REF}

\end{document}